# RF Performance Projections of Graphene FETs vs. Silicon MOSFETs

S. Rodriguez[*], S. Vaziri[*], M. Ostling[*], A. Rusu[*], E. Alarcon[*,#], M.C. Lemme[*1]

[*]KTH Royal Institute of Technology, School of ICT, Kista, Sweden

[#]UPC Universitat Politecnica de Catalunya, Barcelona, Spain

*Abstract*—**A graphene field-effect-transistor (GFET) model calibrated with extracted device parameters and a commercial 65 nm silicon MOSFET model are compared with respect to their radio frequency behavior. GFETs slightly lag behind CMOS in terms of speed despite their higher mobility. This is counterintuitive, but can be explained by the effect of a strongly nonlinear voltage-dependent gate capacitance. GFETs achieve their maximum performance only for narrow ranges of $V_{DS}$ and $I_{DS}$, which must be carefully considered for circuit design. For our parameter set, GFETs require at least $\mu$=3000 cm$^2$ V$^{-1}$ s$^{-1}$ to achieve the same performance as 65nm silicon MOSFETs.**

---

[1] lemme@kth.se

INTRODUCTION

Graphene has attracted enormous research interest in the solid state physics and electronics communities since its experimental discovery in 2004.[1] The unusual electronic band structure of graphene with an energy band gap of 0 eV and a linear dispersion relation leads to charge carriers with extremely high carrier mobilities, with up to $10 \times 10^3$-$15 \times 10^3$ cm$^2$ V$^{-1}$ s$^{-1}$ reported for graphene on SiO$_2$.[2] A high saturation velocity of $>3 \times 10^7$ cm/s has been reported for low carrier densities.[3] Finally, its two-dimensional structure allows the top-down fabrication of graphene field effect transistors (GFETs) using silicon technology.[4] However, due to the absence of a band gap, GFETs are not favorable for logic circuits. On the other hand, under certain DC biasing conditions, GFETs display current saturation, similar to MOS and bipolar devices.[5,6] These saturation regions are of particular interest for analog circuit design as they enable GFETs to be used in amplifier configurations. Furthermore, transit frequencies $f_T$ in excess of 100 GHz suggest GFETs for RF applications.[7,8] Nevertheless, it is still unclear how GFET technology at its present state compares with nanometer CMOS in terms of high frequency circuit design performance metrics.

This letter compares systematically the RF performance of current nanometer CMOS technology and the performance of GFET technology, projected by scaling critical dimensions in a model. The CMOS models used for this comparison belong to a 65nm CMOS commercial process. The GFET model is based on our experimental data to which the models of Meric et al.[5] and Thiele et al.[9] are applied. Key parameters such as minimum sheet carrier concentration $\rho_{sh0}$, Dirac offset voltage $V_{GS\text{-}top0}$, carrier low field mobility $\mu$, and saturation velocity $v_{sat}$ were extracted from experiments and used to fit the model. This Drift-Diffusion model should accurately describe transport at 65nm, as the extracted mobility indicates that fixed charge impurities, charge puddles, substrate roughness and short range scattering centers dominate the transport, which is in line with a channel length-independent saturation velocity down to 130nm reported recently.[10] For shorter devices, the ballisticity of transport will become increasingly dominant, but given the variability in the graphene and device manufacturing processes, a universal description is impossible at the moment. We note that we have not included a model for contact resistance, but only a contact resistance parameter based on.[5] However, RF designs typically allow for ample chip area for low resistance and high current densities in the devices. The GFET model was implemented in Verilog AMS so that it can be simulated using the same circuit design tools and setups as the commercial CMOS process. After comparing the RF performance of the two technologies, a discussion about the impact of $\mu$ on $f_T$ of GFET devices is presented. Finally, a prediction of $f_T$ for technologically viable $\mu$ values is presented.

## EXPERIMENT

Graphene FETs were fabricated on silicon wafers with 285 nm of thermal oxide. Mechanical exfoliation was used to transfer the graphene onto the substrates. After electron beam lithography, 30 nm of tungsten was deposited as source and drain contacts. After evaporation of 30 nm of $SiO_2$, the gate contact was defined by e-beam lithography and lift-off. The inset in Fig. 1a shows an optical micrograph of the GFET, which has a channel length of L = 1 μm and a width of W = 10 μm. The $I_{DS}$-$V_{GS}$ measurement (lines) and fitted model (dotted lines) in Fig. 1a show the typical ambipolar behavior of GFETs and a shift of the Dirac voltage (i.e. the point of minimum conductance) with increasing drain voltage, which can be explained by the influence of the drain voltage on the channel potential.[6] This drain induced Dirac shift (DIDS) is one reason for current saturation in the output characteristics. The extracted parameters after fitting the measured data to the model are: minimum sheet carrier concentration $\rho_{sh0} = 0.7 \times 10^{12}$ cm$^{-2}$, Dirac offset voltage $V_{GS-top0} = 0.5$ V, and carrier low field mobility $\mu = 2500$ cm$^2$ V$^{-1}$ s$^{-1}$. The saturation velocity expression is taken from Thiele et al.[9]:

$$v_{sat} = \frac{\Omega}{(\pi \rho_{sh})^{\frac{1}{2} + \frac{1}{2} V^2(x)}} \qquad (1)$$

where $V(x)$ is the voltage drop at each point in the graphene channel.

With these values extracted from the experimental data, the model allows us to virtually scale the gate length to $L = 65$ nm and the oxide thickness to $T_{OX} = 2.6$ nm. These values correspond to those in the 65 nm CMOS process used for comparison. Fig. 1b shows the simulated drain-source currents $I_{DS}$ as a function of gate-source voltage $V_{GS}$ and drain-source voltage $V_{DS}$ for the scaled GFET. It can be seen that for gate voltages smaller than the Dirac voltage, $I_{DS}$ increases similar to CMOS devices operating in the triode region. As $V_{GS}$ becomes larger (i.e. more positive) than the Dirac voltage, $I_{DS}$ saturates, making possible the design of different amplifying blocks. Beyond $V_{DS} = 1$V, the drain current increases again for increasing $V_{DS}$, as shown in [5] and [9]. Since this is beyond the parameter space for the 65nm CMOS reference used in this work ($V_{DD} = 1.2$ V), we have not plotted this range.

MODEL-BASED PROJECTION OF RF PERFORMANCE METRICS

Typically, the performance of amplifying devices at high frequencies can be compared by looking at the transit frequency $f_T$, which can be expressed as:

$$f_T = \frac{1}{2\pi} \frac{g_m}{C_{tot}} \qquad (2)$$

where $g_m$ and $C_{tot}$ represent the transconductance and total input capacitance. $C_{tot}$ is assumed to be dominated by the gate capacitance $C_G$ of the GFET. The effect of overlap and fringing capacitances between gate-drain, and gate-source terminals are more difficult to estimate since the contact resistances dominate at RF frequencies. When these resistances are very high, these parasitic capacitors can be disregarded for practical purposes. Although these extrinsic parasitic capacitances will reduce somewhat the performance of the device, the intrinsic capacitance $C_G$ still dominates and fundamentally limits the achievable $f_T$.

The value of $C_G$ at each point of the channel is expressed as the series of the top gate oxide capacitance $C_{ox\text{-}top}$ and the quantum capacitance $C_q$. $C_{ox\text{-}top}$ is constant whereas $C_q$ is a function of the voltage drop $V$ at each point in the graphene channel. $V$ is 0V at the source, and $V_{DS}$ at the drain. The capacitance $C_{ox\text{-}back}$ is disregarded since it is short-circuited by the DC source $V_{GS\text{-}back}$. Accordingly, the total value of $C_G$ is obtained by using the following expression:

$$C_G = W \int_0^{V_{DS}} \frac{C_{ox-top} \cdot C_q(V)}{C_{ox-top} + C_q(V)} dV \qquad (3)$$

Fig. 2a shows the simulated values of $C_G$ for the 65 nm GFET. $C_G$ is strongly dependent on $V_{GS}$, with a minimum at the Dirac point. Similar to $g_m$, $C_G$ also depends strongly on $V_{DS}$, which leads to a large variation in $C_G$ magnitude and has a profound impact on the maximum speed of the transistor. This is quite opposite to CMOS transistors, where the overlap capacitance $C_{GD}$ is independent of biasing voltages, and $C_{GS}$ is relatively constant at the saturation region with an approximate value of $2/3 C_{OX} WL$. This situation can also be seen in Fig. 2b where the simulated $f_T$ is plotted against $I_{DS}$ for different $V_{DS}$ voltages. It can be seen that the $f_T$ peaks for $V_{DS}$ of around 210mV and $I_{DS}$ of 1.25mA. We call it the $f_{T,MAX}$ of the device (not to be confused with $f_{MAX}$ where the power gain becomes 1). Larger $V_{DS}$ voltages or $I_{DS}$ currents only reduce the $f_T$. Furthermore, peak performance only happens for narrow ranges of $I_{DS}$, in this case on the order of hundreds of μA. Note that the two peaks for each $V_{DS}$ originate from the fact that we have simulated $f_T$ for both electrons and holes due to the ambipolarity of GFETs. The higher $f_T$ values correspond to $V_{GS} > V_{Dirac}$. Fig. 3a shows $f_T$ simulation results for GFET and CMOS transistors of 10 um width and lengths ranging from 65 nm to 0.25 um. All CMOS transistors are from the same 65 nm CMOS process and are simulated using BSIM 4.1 models. Both GFETS and CMOS transistors are simulated using the same schematic setups and the Cadence Spectre simulation engine. The CMOS devices are biased at the maximum rated voltage specified for this process, $V_{DS} = 1.2$ V. The GFETs are biased at $V_{DS}$ values that provide $f_{T,MAX}$. The first difference that can be seen is that the $f_T$ in CMOS transistors gradually increases from small $I_{DS}$ values whereas the $f_T$ in GFETs can not be defined for $I_{DS}$ values lower than the $I_{DS}$ at the Dirac point. For these $I_{DS}$ values, the GFETs are not suitable as amplifiers. For larger currents, $f_T$ increases sharply, peaks and then decreases. Although the CMOS devices exhibit higher $f_{T,MAX}$ for all gate lengths, this performance is achieved at roughly two times higher current consumption than the $f_{T,MAX}$ of the GFET. At similar current levels of $I_{DS} = 1$ mA, the GFETs perform almost as high as the CMOS devices. Finally, GFETs achieve their best performance only in a very narrow $I_{DS}$ range. This is a critical observation, because it affects the freedom to design for other analog design parameters such as noise and linearity. Even though the GFET mobility in the experimental devices and the model is far superior to the 65 nm MOSFETs, the performance of the GFETs is limited by its lower $g_m$ and parasitics. This is contrary to the common belief that the superior mobility in GFET devices is sufficient to provide better performance than CMOS. The quadratic dependence of $I_{DS}$-$V_{GS}$ in MOS devices seems to provide higher $g_m$ while the intrinsic capacitances are somewhat smaller, therefore resulting in higher $f_T$ values.

As a scaling guideline for future graphene FETs we explored which values of $\mu$ are necessary for GFETs to exceed CMOS performance. Fig. 3b shows simulation results of $f_{T,MAX}$ for a 65 nm GFET transistor when $\mu$ ranges from 500 cm$^2$ V$^{-1}$ s$^{-1}$ to 14×10$^3$ cm$^2$ V$^{-1}$ s$^{-1}$, a reasonable range based on many previous experiments for graphene on SiO$_2$ and well below the intrinsic limit of 40×10$^3$ cm$^2$ V$^{-1}$ s$^{-1}$ induced by phonon scattering.[11] It can be seen that a GFET mobility of $\mu$ = 3000 cm$^2$ V$^{-1}$ s$^{-1}$ is needed to compete with the $f_{T,MAX}$ of 150 GHz obtained in the optimized 65 nm CMOS. Furthermore, if $\mu$ approaches the higher values obtained for graphene on SiO$_2$, then GFETs could perform much better than current nanometer CMOS technologies and approach 1 THz operation. This is an important requirement for the quality of large area graphene films, e.g. fabricated by chemical vapor deposition techniques, where mobility values are typically several thousand cm$^2$ V$^{-1}$ s$^{-1}$ and much lower than in exfoliated graphene.

CONCLUSION

A systematic comparison of RF performance metrics between 65nm GFET and silicon MOSFET models shows that GFETs slightly lag behind in $f_T$ and require at least $\mu$ = 3000 cm$^2$ V$^{-1}$ s$^{-1}$ in order to achieve similar RF performance. While a strongly nonlinear voltage-dependent gate capacitance inherently limits performance, other parasitics such as contact resistance are expected to be optimized as GFET process technology improves. Finally, this letter quantifies the $\mu$ values, which would allow future GFETs to match and exceed CMOS, potentially up to THz operation.

ACKNOWLEDGEMENT

The authors gratefully acknowledge support through an Advanced Investigator Grant (OSIRIS, No. 228229) and a Starting Grant (InteGraDe, No. 307311) from the European Research Council.

FIGURES

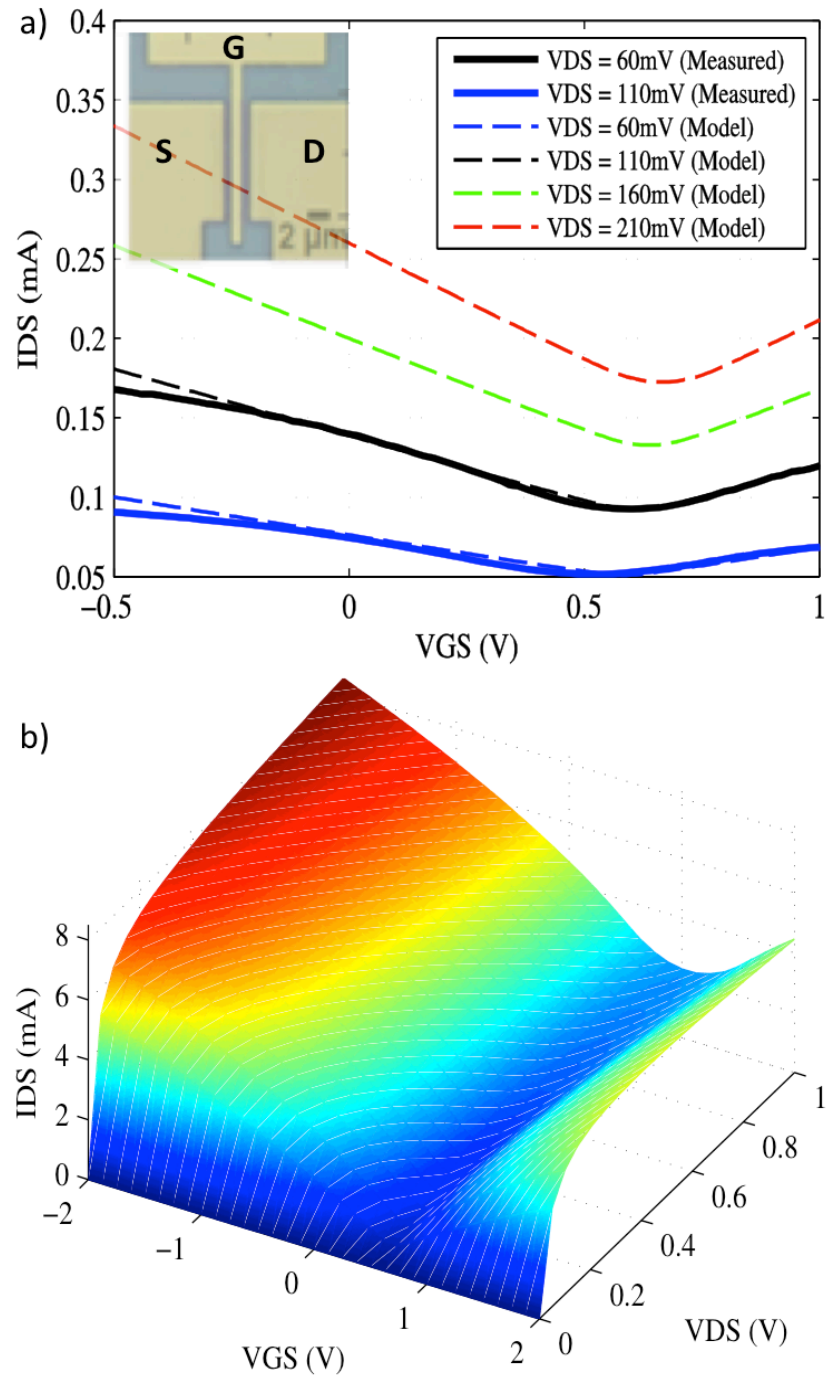

Figure 1: a) Measured (solid lines) and modeled (dashed lines) transfer characteristics of a GFET with a gate width of W=10 μm and a gate length of L = 1 μm used as the basis for this work. Inset: Optical microscope image of the device (false color). b) Modeled drain current $I_{DS}$ for different $V_{GS}$ and $V_{DS}$ bias conditions for virtually scaled GFET with L = 65 nm and W = 10 μm.

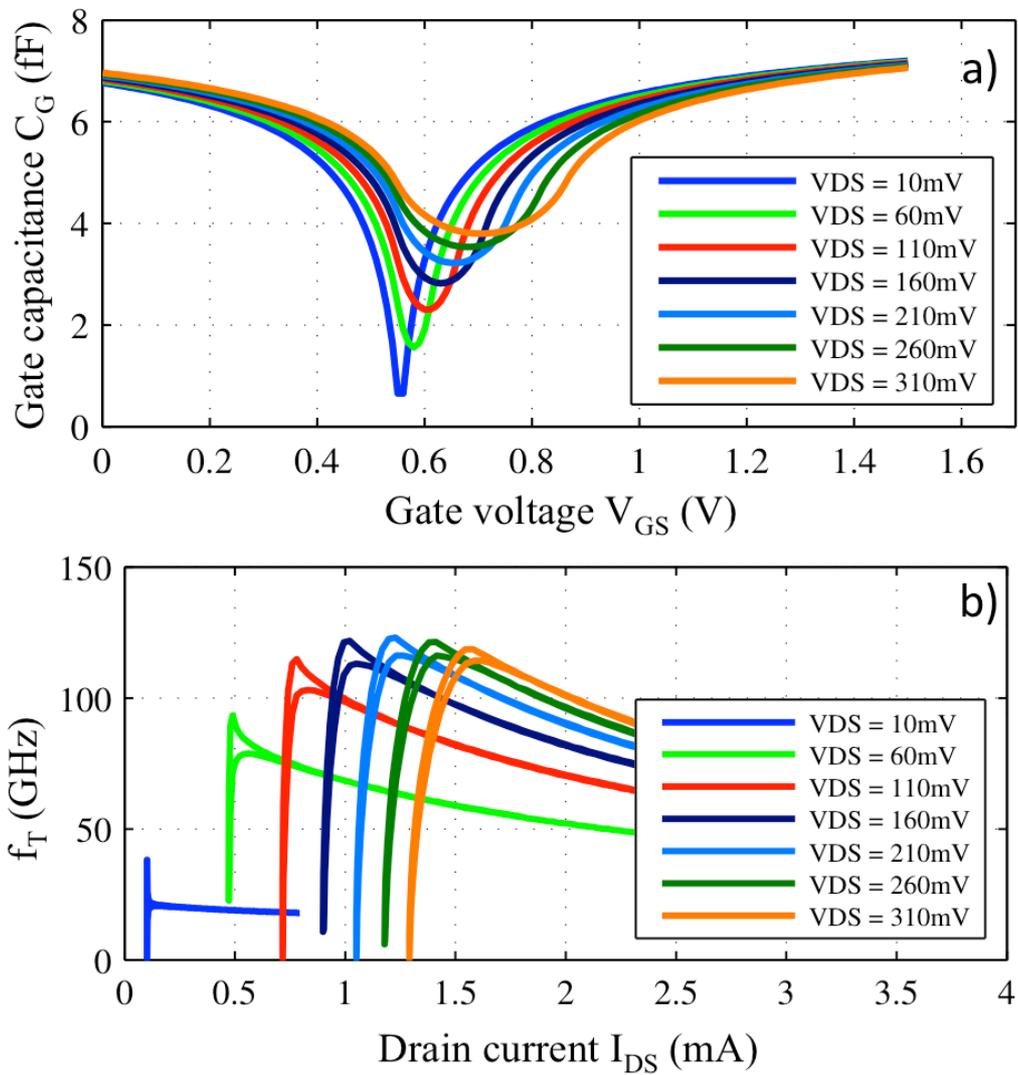

Figure 2: a) Top gate capacitance $C_G$ vs. gate voltage $V_{GS}$ for various drain bias voltages $V_{DS}$. b) Cut-off frequency $f_T$ vs. drain current $I_{DS}$ for various drain bias voltages $V_{DS}$. (L = 65 nm and W = 10 µm).

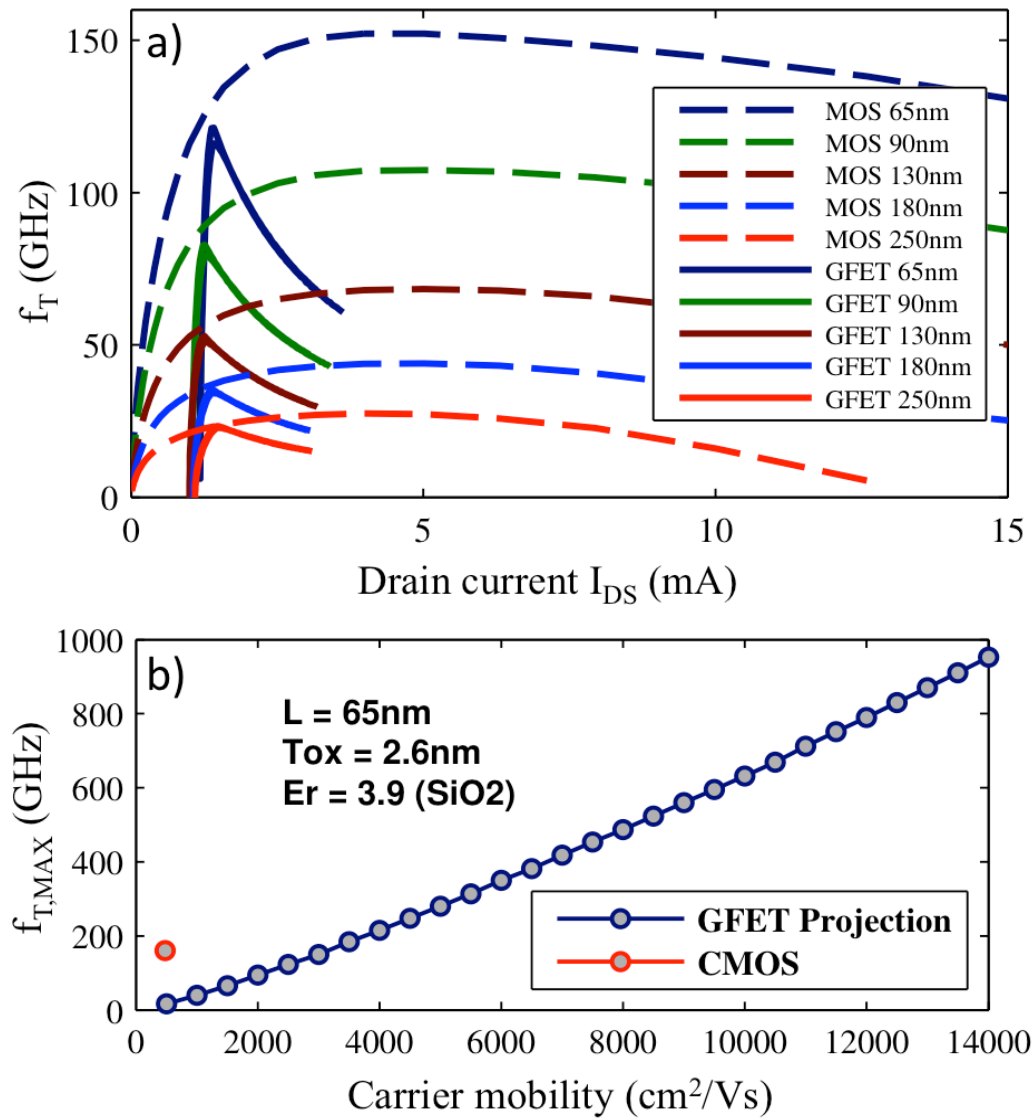

Figure 3: a) Simulated cut-off frequency $f_T$ vs. drain current $I_{DS}$ for various Si-MOSFETs and GFETs with a fixed gate width of 10 µm and various gate lengths. b) Simulated maximum cut-off frequency $f_{T,MAX}$ vs. mobility µ for GFETs with a gate length of L = 65 nm, gate oxide thickness of $T_{OX}$ = 2.6 nm (SiO$_2$; $\varepsilon_r$ = 3.9).